# Minimal Filtering Algorithms for Convolutional Neural Networks

Aleksandr Cariow [1,*] and Galina Cariowa[1]

[1] West Pomeranian University of Technology Szczecin, Żołnierska 52,
71-210 Szczecin Poland

* Correspondence: acariow@wi.zut.edu.pl

**Abstract:** In this paper, we present several resource-efficient algorithmic solutions regarding the fully parallel hardware implementation of the basic filtering operation performed in the convolutional layers of convolution neural networks. In fact, these basic operations calculate two inner products of neighboring vectors formed by a sliding time window from the current data stream with an impulse response of the *M*-tap finite impulse response filter. We used Winograd's minimal filtering trick and applied it to develop fully parallel hardware-oriented algorithms for implementing the basic filtering operation for $M$= 3, 5, 7, 9, and 11. A fully parallel hardware implementation of the proposed algorithms in each case gives approximately 30% savings in the number of embedded multipliers compared to a fully parallel hardware implementation of the naive calculation methods.

## 1. Introduction

Today, artificial intelligence, deep learning and neural networks are powerful and incredibly effective machine learning methods used to solve many scientific and practical problems. Applications of deep neural networks for machine learning are diverse and rapidly developing, covering various areas of basic sciences, technologies and the real world [1, 2]. Among the various types of deep neural networks, convolutional neural networks (CNNs) are most widely used [3]. Although there are many optimizing methods to speed up CNN-based digital signal and image processing algorithms, it is still difficult to implement these algorithms in real-time low-power systems. The main and most time-consuming operations in CNN are two-dimensional convolution operations. To speed up convolution computation, various algorithmic methods have been proposed [4-13]. The most common approach for efficient convolution implementation is the Fast Fourier Transform (FFT) algorithm [4-5]. The FFT-based convolution method is traditionally used for large length finite impulse response (FIR) filters, but modern CNNs use predominantly small length FIR filters. In this situation one of the most effective algorithms used in the computation of a small-length two-dimensional convolution is the Winograd's minimal filtering algorithm, which is most intensively used in recent time [6-13]. The algorithm computes linear convolution over small tiles with minimal complexity, which makes it more effective with small filters and small batch sizes. In fact, this algorithm calculates two inner products of neighboring vectors formed by a sliding time window from the current data stream with an impulse response of the 3-tap FIR filter.

CNN contains several kinds of layers. However, the name of the convolutional neural network itself suggests that the convolutional layers are dominant in this type of network. In CNN, convolutional layers are the most computationally intensive, since in a typical implementation they occupied more than 90% of the CNN execution time [14]. In turn, convolution itself requires performing a large number of arithmetic operations. In many cases, convolution is performed on terabytes or petabytes of data, so even insignificant improvement can significantly reduce the computation time. That is why developers of such type networks seek and design efficient ways of implementing convolution using the smallest possible number of arithmetic operations. Especially, algorithm developers try to minimize the number of multiplications since this operation is more complex than addition. Despite the fact that the execution time of addition and multiplication in modern computers is supposedly comparable, nevertheless, multiplication requires more manipulations with operands, therefore its implementation requires more time and effort than expected. As a result of the multiplication of two $n$-bit operands, a $2n$-bit product is obtained. This is why in all fixed-point digital signal processing (DSP) units the product register and the accumulator are double the widths of all other registers. However, in such a case, two-time access to memory during both writing and reading is required. This increases the actual multiplication time. For example, 32-bit integer multiplication on GPU takes 16 clock cycles. Floating-point multiplication operations require even rather more complicated housekeeping. Therefore, the statement that in modern processors the multiplication operation takes the same time as the addition is somewhat exaggerated.

Another way to solve this problem is to take advantage of the massive parallelism offered by graphic processing units (GPUs), application-specific integrated circuit (ASIC) and field programmable gate array (FPGA) devices to implement a large amount of internal parallelism demonstrated by CNN-based algorithms [14-30]. GPUs are the most popular and widely used accelerators for improving training and classification processes at CNN [15-17]. This is due to their high performance when performing matrix operations [18]. However, GPU accelerators consume a large amount of energy and therefore their use in CNN-based applications implemented in on-board battery-powered mobile devices is becoming a problem. ASIC and FPGA are the preferred acceleration platforms on-board CNN due to their promising performance and high energy efficiency. They can also achieve high performance, but with significantly lower power consumption [19-30]. In addition, most modern high-performance FPGA targets contain a number of integrated hardware multipliers. Thus, instead of mapping a multiplier into several logic gates, dedicated multipliers provided on the FPGA fabric can be used. So, all multiplications involved in the implementation of the fully parallel algorithm can be efficiently implemented using these embedded multipliers. However, their number may simply be insufficient to meet the requirements of a fully parallel implementation of the algorithm. If multiplications are implemented using hardwired multipliers within the target FPGA, this dramatically limits the complexity of the CNN that can be implemented. For example, the second layer of the LeNet5 network requires 2400 multipliers [31]. This number largely exceeds the number of multipliers provided by many FPGAs and, especially by embedded devices. The designer uses hardwired multipliers to implement multiplication operations until the implemented computing unit occupies all the embedded hardwired multipliers. If the FPGA target runs out of embedded multipliers, the designer uses generic logic gates instead, and the multiplication implementation becomes expensive in terms of FPGA resource usage. In some cases, therefore, available logic has to be exploited to implement multipliers, seriously restricting the maximum number of real multiplications that can be implemented in parallel on a target device. This will lead to significant difficulties during the implementation of the computation unit. Thus, the problem of minimizing the multiplications in the development of the parallel hardware-oriented algorithms for convolutional neural networks regardless of which platforms they will be implemented remains relevant. Next, we consider a number of algorithmic solutions that contribute to the solution of this problem.

**2. Preliminary Remarks**

The main operation of convolutional neural networks is an inner product of a vector, formed by a sliding time window from the current data stream with an impulse response of the $M$-tap finite impulse response (FIR) filter. In the most general case, the procedure for calculating convolution elements can be represented as follows:

$$y_j = \sum_{i=0}^{M-1} x_{i+j} w_i \tag{1}$$

$$i = 0,1,...,M-1, \quad j = 0,1,...,N-M+1$$

where $N$ is a length of current data stream, $\{x_{i+j}\}$ are the elements of the current data stream, $\{w_i\}$ are the coefficients of the impulse response of the FIR filter, which are constants.

For example, a direct application of two consecutive steps of a 3-tap FIR filter with coefficients $\{w_0, w_1, w_2\}$ to a set of four elements $\{x_0, x_1, x_2, x_3\}$ requires 4 additions and 6 multiplications:

$$y_0 = x_0 w_0 + x_1 w_1 + x_2 w_2, \quad y_1 = x_1 w_0 + x_2 w_1 + x_3 w_2 \tag{2}$$

S. Winograd came up with a tricky way to reduce the number of multiplications during calculating expression (2):

$$\mu_1 = (x_0 - x_2)w_0, \quad \mu_2 = (x_1 + x_2)\frac{w_0 + w_1 + w_2}{2}, \quad \mu_3 = (x_2 - x_1)\frac{w_0 - w_1 + w_2}{2}, \quad \mu_4 = (x_1 - x_3)w_2,$$

$$y_0 = \mu_1 + \mu_2 + \mu_3, \quad y_1 = \mu_2 - \mu_3 - \mu_4.$$

This trick was called the minimal filtering algorithm [7]. The values $(w_0 + w_1 + w_2)/2$ and $(w_0 - w_1 + w_2)/2$ can be calculated in advance, then this method requires 4 multiplications and 8 additions, which is equal to number of arithmetical operations in the direct method. Since multiplication is a much more complicated operation than addition, the Winograd's minimal filtering algorithm is more efficient than the direct method of computation.

The above expressions exhaustively describe the entire set of mathematical operations necessary to perform the calculations. But, strictly speaking, they are not an algorithm, because they do not reveal the sequence of calculations. In addition, convolutional neural networks use FIR filters with a longer impulse response, for which minimal filtering algorithms have not yet been developed.

Considering the above, the goal of this article is to develop and describe minimal filtering algorithms for $M$ = 3, 5, 7, 9, 11.

First, we define the basic operation of CTT-filtering as an application of two consecutive steps of an $M$-tap FIR filter with coefficients $\{w_0, w_1, ..., w_{M-1}\}$ to a set of elements $\{x_0, x_1, ..., x_M\}$. Figure 1 clarifies the essence of what was said.

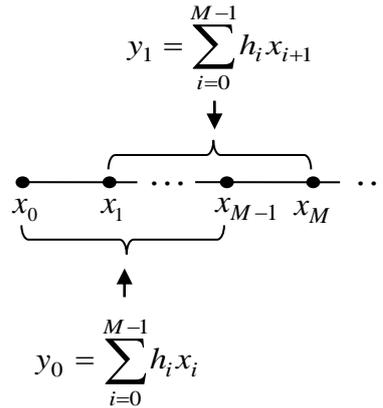

**Figure 1.** Illustration of the organization of calculations in accordance with the basic filtering operation.

A more compactly introduced operation can be represented in the form of a vector-matrix product:

$$\mathbf{y}_2 = \mathbf{F}_{2 \times M} \mathbf{w}_M \tag{3}$$

where

$$\mathbf{F}_{2\times M} = \begin{bmatrix} x_0 & x_1 & \cdots & x_{M-1} \\ x_1 & x_2 & \cdots & x_M \end{bmatrix}, \quad \mathbf{y}_2^{(M)} = [y_0^{(M)}, y_1^{(M)}]^T, \quad \mathbf{w}_M = [w_0^{(M)}, w_1^{(M)}, \ldots, w_{M-1}^{(M)}]^T.$$

(Please note, that hereinafter, the superscript (*M*) will denote quantities related to the basic operation of minimal filtering with an *M*-tap filter).

Next, we present minimal filtering algorithms using a Winograd's trick for 3-tap FIR filter. The developed algorithms are distinguished by a reduced number of multiplications, which makes them suitable for fully parallel hardware implementation.

## 3. Minimal filtering algorithms

*3.1. Algorithm 1, M=3.*

Let $\mathbf{x}_4 = [x_0, x_1, x_2, x_3]^T$ be a vector that represents the input data set, $\mathbf{w}_3 = [w_0^{(3)}, w_1^{(3)}, w_2^{(3)}]^T$ be a vector that contains the coefficients of the impulse response of 3-tap FIR filter, and $\mathbf{y}_2^{(3)} = [y_0^{(3)}, y_1^{(3)}]^T$ be a vector describing the results of using a 3-tap FIR filter. Then, a fully parallel algorithm for computation $\mathbf{y}_2^{(3)}$ using Winograd's minimal filtering method can be written with the help of following matrix-vector calculating procedure:

$$\mathbf{y}_2^{(3)} = \mathbf{A}_{2\times 4}^{(3)} \mathbf{D}_4^{(3)} \mathbf{A}_4^{(3)} \mathbf{x}_4 \qquad (4)$$

where

$$\mathbf{A}_4^{(3)} = \begin{bmatrix} 1 & -1 \\ 1 & 1 \\ -1 & 1 \\ 1 & -1 \end{bmatrix}, \quad \mathbf{A}_{2\times 4}^{(3)} = \begin{bmatrix} 1 & 1 & 1 & 0 \\ 0 & 1 & -1 & -1 \end{bmatrix},$$

and

$$\mathbf{D}_4^{(3)} = diag(s_0^{(3)}, s_1^{(3)}, s_2^{(3)}, s_3^{(3)}),$$

$$s_0 = w_0^{(3)}, \quad s_1^{(3)} = (w_0^{(3)} + w_1^{(3)} + w_2^{(3)})/2, \quad s_2^{(3)} = (w_0^{(3)} - w_1^{(3)} + w_2^{(3)})/2, \quad s_3^{(3)} = w_2^{(3)}.$$

Figure 2 shows a data flow diagram of the proposed algorithm for the implementation of minimal filtering basic operation for 3-tap FIR filter. In this paper, data flow diagrams are oriented from left to right and straight lines in the figures denote the data transfer operations. The circles in these figures show the operation of multiplication by a number inscribed inside a circle. The points where the lines converge indicate the summation, the dashed lines indicate data transfer operations with a simultaneous change of sign. We use the usual lines without arrows on purpose, so as not to clutter the picture. In order to simplify, we also removed the superscripts of the variables in all the figures, since it is obvious from the figures what vector sizes we are dealing with in each case.

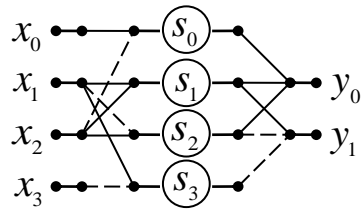

**Figure 2.** Illustration of the organization of calculations in accordance with the basic filtering operation, *M*=3.

*3.2. Algorithm 2, M=5.*

Let $\mathbf{x}_6 = [x_0, x_1, ..., x_5]^T$ be a vector that represents the input data set, $\mathbf{w}_5 = [w_0^{(5)}, w_1^{(5)}, ..., w_4^{(5)}]^T$ be a vector that contains the coefficients of the impulse response of 5-tap FIR filter, and $\mathbf{y}_2^{(5)} = [y_0^{(5)}, y_1^{(5)}]^T$ be a vector describing the results of using a 5-tap FIR filter. Then, a fully parallel minimal filtering algorithm for computation $\mathbf{y}_2^{(5)}$ can be written with the help of following matrix-vector calculating procedure:

$$\mathbf{y}_2^{(5)} = \mathbf{A}_{2\times 7}^{(5)} \mathbf{D}_7^{(5)} \mathbf{A}_{7\times 6}^{(5)} \mathbf{x}_6^{(5)} \tag{5}$$

where

$$\mathbf{A}_{7\times 6}^{(5)} = \begin{bmatrix} 1 & -1 & & & & \\ & 1 & 1 & & & \\ & -1 & 1 & & & \\ & & 1 & & -1 & \\ & & & 1 & -1 & \\ & & & & 1 & \\ & & & & -1 & 1 \end{bmatrix}, \quad \mathbf{A}_{2\times 7}^{(5)} = \begin{bmatrix} 1 & 1 & 1 & & 1 & 1 & \\ & 1 & -1 & -1 & & 1 & 1 \end{bmatrix}, \quad \mathbf{D}_7^{(5)} = diag(s_0^{(5)}, s_1^{(5)}, ..., s_6^{(5)}),$$

$s_0^{(5)} = w_0^{(5)}$, $s_1^{(5)} = (w_0^{(5)} + w_1^{(5)} + w_2^{(5)})/2$, $s_2^{(5)} = (w_0^{(5)} - w_1^{(5)} + w_2^{(5)})/2$, $s_3^{(5)} = w_2^{(5)}$, $s_4^{(5)} = w_3^{(5)}$,

$s_5^{(5)} = w_3^{(5)} + w_4^{(5)}$, $s_6^{(5)} = w_4^{(5)}$.

Figure 3 shows a data flow diagram of the proposed algorithm for the implementation of minimal filtering basic operation for 5-tap FIR filter.

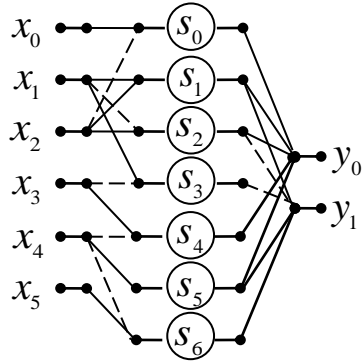

**Figure** 3. Illustration of the organization of calculations in accordance with the basic filtering operation, *M*=5.

*3.3. Algorithm 3, M=7.*

Let $\mathbf{x}_9 = [x_0, x_1, ..., x_8]^T$ be a vector that represents the input data set, $\mathbf{w}_7 = [w_0^{(7)}, w_1^{(7)}, ..., w_6^{(7)}]^T$ be a vector that contains the coefficients of the impulse response of 7-tap FIR filter, and $\mathbf{y}_2^{(7)} = [y_0^{(7)}, y_1^{(7)}]^T$ be a vector describing the results of using a 7-tap FIR filter. Then, a fully parallel minimal filtering algorithm for computation $\mathbf{y}_2^{(7)}$ can be written with the help of following matrix-vector calculating procedure:

$$\mathbf{y}_2^{(7)} = \mathbf{A}_{2\times 6}^{(7)} \mathbf{A}_{6\times 10}^{(7)} \mathbf{D}_{10}^{(7)} \mathbf{A}_{10\times 8}^{(7)} \mathbf{x}_8^{(7)} \tag{6}$$

where

$$\mathbf{A}_{10\times 8}^{(7)} = \begin{bmatrix} 1 & -1 & & & & & & \\ 1 & 1 & & & & & & \\ -1 & 1 & & & \mathbf{0}_{5\times 4} & & \\ 1 & -1 & & & & & & \\ & & 1 & & & & & \\ \hline & & & 1 & & & & \\ & & & & 1 & -1 & & \\ & \mathbf{0}_{5\times 4} & & & 1 & 1 & \\ & & & & -1 & 1 & \\ & & & & 1 & -1 \end{bmatrix}, \quad \mathbf{A}_{6\times 10}^{(7)} = \begin{bmatrix} 1 & 1 & 1 & \mathbf{0}_2 & & \mathbf{0}_{2\times 4} \\ 1 & -1 & -1 & & & \\ \hline \mathbf{0}_{2\times 4} & & 1 & & \mathbf{0}_{2\times 4} \\ & & & 1 & & \\ \hline \mathbf{0}_{2\times 4} & & \mathbf{0}_2 & 1 & 1 & 1 \\ & & & & 1 & -1 & -1 \end{bmatrix},$$

$$\mathbf{A}_{2\times 6}^{(7)} = \begin{bmatrix} 1 & & 1 & & 1 & \\ & 1 & & 1 & & 1 \end{bmatrix}, \quad \mathbf{D}_{10}^{(7)} = diag(s_0^{(7)} s_1^{(7)},...,s_9^{(7)}),$$

$s_0^{(7)} = w_0^{(7)}$, $s_1^{(7)} = (w_0^{(7)} + w_1^{(7)} + w_2^{(7)})/2$, $s_2^{(7)} = (w_0^{(7)} - w_1^{(7)} + w_2^{(7)})/2$, $s_3^{(7)} = w_2^{(7)}$, $s_4^{(7)} = s_5^{(7)} = w_3^{(7)}$,

$s_6^{(7)} = w_4^{(7)}$, $s_7^{(7)} = (w_4^{(7)} + w_5^{(7)} + w_6^{(7)})/2$, $s_8^{(7)} = (w_4^{(7)} - w_5^{(7)} + w_6^{(7)})/2$, $s_9^{(7)} = w_6^{(7)}$.

Figure 4 shows a data flow diagram of the proposed algorithm for the implementation of minimal filtering basic operation for 7-tap FIR filter.

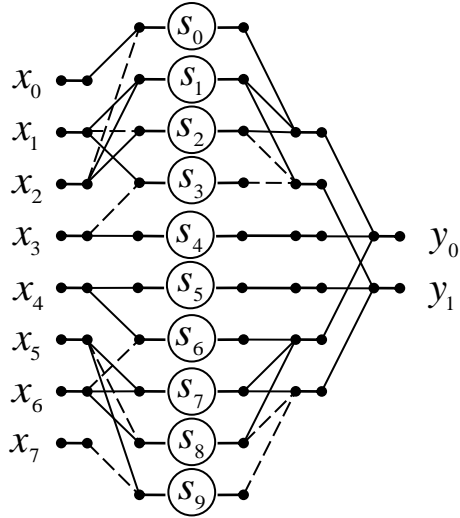

**Figure 4.** Illustration of the organization of calculations in accordance with the basic filtering operation, M=7.

*3.4. Algorithm 4, M=9*

Let $\mathbf{x}_{10} = [x_0, x_1,...,x_9]^T$ be a vector that represents the input data set, $\mathbf{w}_9 = [w_0^{(9)}, w_1^{(9)},...,w_8^{(9)}]^T$ be a vector that contains the coefficients of the impulse response of 9-tap FIR filter, and $\mathbf{y}_2^{(9)} = [y_0^{(9)}, y_1^{(9)}]^T$ be a vector describing the results of using a 9-tap FIR filter. Then, a fully parallel minimal filtering algorithm for computation $\mathbf{y}_2^{(9)}$ can be written with the help of following matrix-vector calculating procedure:

$$\mathbf{y}_2^{(9)} = \mathbf{A}_{2\times 6}^{(9)} \mathbf{A}_{6\times 12}^{(9)} \mathbf{D}_{12}^{(9)} \mathbf{A}_{12\times 10}^{(9)} \mathbf{x}_{10}^{(9)} \qquad (7)$$

where

$$\mathbf{A}_{12\times 10}^{(9)} = \begin{bmatrix} 1 & -1 & & & & & & & & \\ 1 & 1 & & & & & & & & \\ -1 & 1 & & & \mathbf{0}_{4\times 3} & & & & \\ 1 & -1 & & & & & & & & \\ \hdashline & & 1 & -1 & & & & & & \\ & & 1 & 1 & & & & & & \\ \mathbf{0}_{4\times 3} & & -1 & 1 & & & & & \\ & & 1 & -1 & & & & & \\ \hdashline & & & & & 1 & -1 & & & \\ & & & & & 1 & 1 & & & \\ \mathbf{0}_{4\times 3} & \mathbf{0}_{4\times 3} & & & -1 & 1 & & \\ & & & & & 1 & -1 \end{bmatrix},$$

$$\mathbf{A}_{6\times 12}^{(9)} = \begin{bmatrix} 1 & 1 & 1 & & & & & & & \\ & 1 & -1 & -1 & \mathbf{0}_{2\times 4} & & \mathbf{0}_{2\times 4} & \\ \hdashline \mathbf{0}_{2\times 4} & & 1 & 1 & 1 & & \mathbf{0}_{2\times 4} & \\ & & & 1 & -1 & -1 & & \\ \hdashline \mathbf{0}_{2\times 4} & & \mathbf{0}_{2\times 4} & & 1 & 1 & 1 \\ & & & & & 1 & -1 & -1 \end{bmatrix}, \quad \mathbf{A}_{2\times 6}^{(9)} = \mathbf{A}_{2\times 6}^{(9)} = \begin{bmatrix} 1 & | & 1 & | & 1 & \\ \hdashline & 1 & | & 1 & | & 1 \end{bmatrix},$$

$\mathbf{D}_{12}^{(9)} = diag(s_0^{(9)} s_1^{(9)},...,s_{11}^{(9)})$, $s_0^{(9)} = w_0^{(9)}$, $s_1^{(9)} = (w_0^{(9)} + w_1^{(9)} + w_2^{(9)})/2$, $s_2^{(9)} = (w_0^{(9)} - w_1^{(9)} + w_2^{(9)})/2$,

$s_3^{(9)} = w_2^{(9)}$, $s_4^{(9)} = w_3^{(9)}$, $s_5^{(9)} = (w_3^{(9)} + w_4^{(9)} + w_5^{(9)})/2$, $s_6^{(9)} = (w_3^{(9)} - w_4^{(9)} + w_5^{(9)})/2$, $s_7^{(9)} = w_5^{(9)}$,

$s_8^{(9)} = w_6^{(9)}$, $s_9^{(9)} = (w_6^{(9)} + w_7^{(9)} + w_8^{(9)})/2$, $s_{10}^{(9)} = (w_6^{(9)} - w_7^{(9)} + w_8^{(9)})/2$, $s_{11}^{(9)} = w_8^{(9)}$.

Figure 5 shows a data flow diagram of the proposed algorithm for the implementation of minimal filtering basic operation for 9-tap FIR filter.

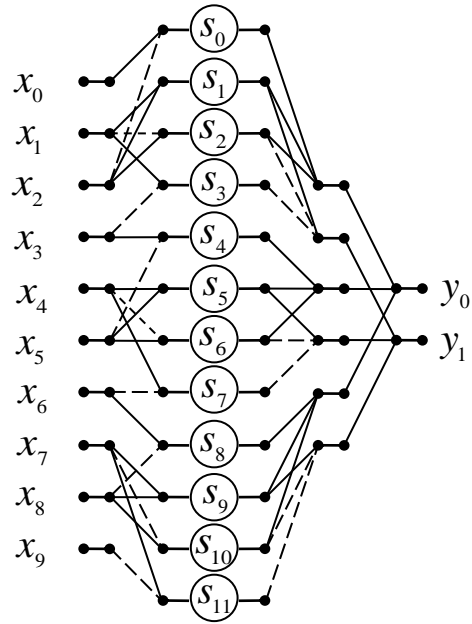

**Figure** 5. Illustration of the organization of calculations in accordance with the basic filtering operation, *M*=9.

*3.5. Algorithm 5, M=11.*

Let $\mathbf{x}_{12}=[x_0,x_1,...,x_{11}]^T$ be a vector that represents the input data set, $\mathbf{w}_{11}=[w_0^{(11)},w_1^{(11)},...,w_{10}^{(11)}]^T$ be a vector that contains the coefficients of impulse response of 11-tap FIR filter, and $\mathbf{y}_2^{(11)}=[y_0^{(11)},y_1^{(11)}]^T$ be a vector describing the results of using a 9-tap FIR filter.

Then, a fully parallel minimal filtering algorithm for computation $\mathbf{y}_2^{(11)}$ can be written with the help of following matrix-vector calculating procedure:

$$\mathbf{y}_2^{(11)} = \mathbf{A}_{2\times 8}^{(11)} \mathbf{A}_{8\times 15}^{(11)} \mathbf{D}_{15}^{(11)} \mathbf{A}_{15\times 12}^{(11)} \mathbf{x}_{12}^{(11)} \quad (8)$$

where

$$\mathbf{A}_{15\times 12}^{(11)} = \begin{bmatrix}
1 & -1 & & & & & & & & & & \\
 & 1 & 1 & & & & & & & & & \\
 & -1 & 1 & & & \mathbf{0}_{4\times 3} & & & \mathbf{0}_{4\times 3} & & \\
 & 1 & -1 & & & & & & & & & \\
\hline
 & & & 1 & -1 & & & & & & & \\
 & & & & 1 & 1 & & & & & & \\
\mathbf{0}_{4\times 3} & & & -1 & 1 & & & & \mathbf{0}_{4\times 3} & & \\
 & & & 1 & -1 & & & & & & & \\
\hline
 & & & & & & 1 & -1 & & & & \\
 & & & & & & & 1 & 1 & & & \\
\mathbf{0}_{4\times 3} & & & \mathbf{0}_{4\times 3} & & -1 & 1 & & & & \\
 & & & & & & 1 & -1 & & & & \\
\hline
 & & & & & & & & & 1 & -1 & \\
\mathbf{0}_3 & & & \mathbf{0}_3 & & \mathbf{0}_3 & & & & 1 & & \\
 & & & & & & & & & -1 & 1 & 
\end{bmatrix},$$

$$\mathbf{A}_{8\times 15}^{(11)} = \begin{bmatrix}
1 & 1 & 1 & & & & \mathbf{0}_{2\times 4} & & \mathbf{0}_{2\times 4} & & \mathbf{0}_{2\times 3} \\
1 & -1 & -1 & & & & & & & & \\
\hline
 & & & 1 & 1 & 1 & & & & & \\
\mathbf{0}_{2\times 4} & & 1 & -1 & -1 & & \mathbf{0}_{2\times 4} & & \mathbf{0}_{2\times 3} \\
\hline
 & & & & & & 1 & 1 & 1 & & \\
\mathbf{0}_{2\times 4} & & \mathbf{0}_{2\times 4} & & 1 & -1 & -1 & & \mathbf{0}_{2\times 3} \\
\hline
 & & & & & & & & & 1 & 1 \\
\mathbf{0}_{2\times 4} & & \mathbf{0}_{2\times 4} & & \mathbf{0}_{2\times 4} & & & 1 & 1
\end{bmatrix},$$

$$\mathbf{A}_{2\times 8} = \mathbf{1}_{1\times 4} \otimes \mathbf{I}_2 = \begin{bmatrix} 1 & & 1 & & 1 & & 1 & \\ & 1 & & 1 & & 1 & & 1 \end{bmatrix}, \quad \mathbf{D}_{15}^{(11)} = diag(s_0^{(11)} s_1^{(11)},...,s_{14}^{(11)}),$$

$s_0^{(11)} = w_0^{(11)}$, $s_1^{(9)} = (w_0^{(11)} + w_1^{(11)} + w_2^{(11)})/2$, $s_2^{(11)} = (w_0^{(11)} - w_1^{(11)} + w_2^{(11)})/2$, $s_3^{(11)} = w_2^{(11)}$, $s_4^{(11)} = w_3^{(11)}$,

$s_5^{(11)} = (w_3^{(11)} + w_4^{(11)} + w_5^{(11)})/2$, $s_6^{(11)} = (w_3^{(11)} - w_4^{(11)} + w_5^{(11)})/2$, $s_7^{(11)} = w_5^{(11)}$, $s_8^{(11)} = w_6^{(11)}$,

$s_9^{(11)} = (w_6^{(11)} + w_7^{(11)} + w_8^{(11)})/2$, $s_{10}^{(11)} = (w_6^{(11)} - w_7^{(11)} + w_8^{(11)})/2$, $s_{11}^{(11)} = w_8^{(11)}$, $s_{12}^{(11)} = w_9^{(11)}$,

$$s_{13}^{(11)} = w_9^{(11)} + w_{10}^{(11)}, \quad s_{14}^{(11)} = w_{10}^{(11)}.$$

Fig. 6 shows a data flow diagram of the proposed algorithm for the implementation of minimal filtering basic operation for 11-tap FIR filter.

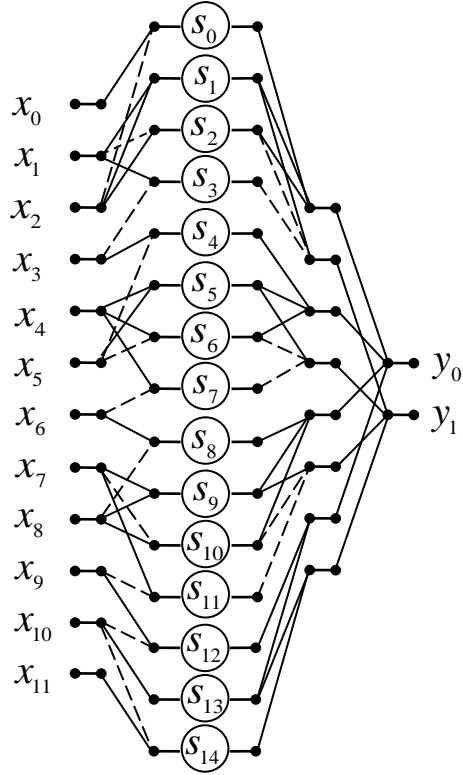

**Figure 6.** Illustration of the organization of calculations in accordance with the basic filtering operation, $M$=11.

**4. Implementation Complexity**

Since the lengths of the input sequences are relatively small, and the data flow diagrams representing the organization of the computation process are fairly simple, it is easy to estimate the implementation complexity of the proposed solutions. Table 1 shows estimates of the number of arithmetic blocks for the fully parallel implementation of the short lengths CNN-minimal filtering algorithms.

**Table 1.** Implementation complexities of naive and proposed solutions.

| Size $M$ | Numbers of arithmetic blocks | | | | | | |
|---|---|---|---|---|---|---|---|
| | Naive method | | Proposed algorithm | | | | |
| | multipliers | $M$-input adders | multipliers | 2-input adders | 3-input adders | 4-input adders | 5-input adders |
| 3 | 6 | 2 | 4 | 4 | 2 | – | 2 |
| 5 | 10 | 2 | 7 | 6 | – | – | – |
| 7 | 14 | 2 | 10 | 8 | 6 | – | – |
| 9 | 18 | 2 | 12 | 12 | 8 | – | – |
| 11 | 22 | 2 | 15 | 16 | 6 | 2 | – |

As you can see, the implementation of the proposed algorithms requires fewer multipliers than the implementation based on naive methods of performing the filtering operations. Reducing the number of multipliers is especially important in the design of specialized VLSI fully parallel processors because minimizing the number of necessary multipliers also reduces the power dissipation and lowers the cost implementation of the entire system being implemented. This is because the hardware multiplier is a more complex unit than the adder and occupies much more of the chip area than the adder. It is proved that the implementation complexity of a hardwired multiplier grows quadratically with operand size, while the hardware complexity of a binary adder increases linearly with operand size [32]. Therefore, a reduction in the number of multipliers, even at the cost of a small increase in the number of adders, has a significant role in the hardware implementation of the algorithm.

## 5. Conclusion

In this paper, we analyzed possibilities to reduce the multiplicative complexity of calculating basic filtering operations for small length impulse responses of the $M$-tap FIR filters, that used in convolution neural networks. We also synthesized new algorithms for implementing these operations for $M$ = 3, 5, 7, 9, 11. Using these algorithms reduces the computational complexity of basic filtering operation, thus reducing its hardware implementation complexity. In addition, as can be seen from Figures, the proposed algorithms have a pronounced parallel modular structure. This simplifies the mapping of the algorithms into an ASIC structure and unifies its implementation in FPGAs. Thus, the acceleration of computations during the implementation of these algorithms can also be achieved due to the parallelization of the computation processes.